\documentclass[letter,scriptaddress,twocolumn, prl,showkeys,showpacs,superscriptaddress]{revtex4-1}
	\usepackage{amsmath}
	\usepackage{makeidx}
	\usepackage{amsfonts}
	\usepackage{mathrsfs}
	\usepackage{graphicx, xcolor}  
	\usepackage{dcolumn}   
	\usepackage{bm}        
	\usepackage{amssymb}   
	\usepackage[ansinew]{inputenc}
	\usepackage[usenames,dvipsnames]{pstricks}
	\usepackage{subfigure}
	\usepackage{epstopdf}
	\usepackage{pst-grad} 
	\usepackage{pst-plot} 
	\usepackage[colorlinks,hyperindex]{hyperref}
	\hypersetup
	{
		colorlinks,%
		citecolor=black,%
		linkcolor=black,%
		urlcolor=black,%
	}
	
\begin{document}


\title{Scale-free channeling patterns  near the onset of erosion of sheared granular beds}
\author{Pascale Aussillous\thanks{pascale.aussillous@univ-amu.fr}}\affiliation{Aix-Marseille Universit\'e, CNRS, IUSTI UMR 7343, 13453 Marseille, France} 
\author{Zhenhai Zou}\affiliation{Aix-Marseille Universit\'e, CNRS, IUSTI UMR 7343, 13453 Marseille, France} 
\author{\'Elisabeth Guazzelli\thanks{elisabeth.guazzelli@univ-amu.fr}}
\affiliation{Aix-Marseille Universit\'e, CNRS, IUSTI UMR 7343, 13453 Marseille, France} 
\author{Le Yan\thanks{lyan@kitp.ucsb.edu}}\affiliation{Kavli Institute for Theoretical Physics, University of California, Santa Barbara, CA 93106, USA} 
\author{Matthieu Wyart\thanks{matthieu.wyart@epfl.ch}}
\affiliation{Institute of Physics, EPFL, CH-1015 Lausanne, Switzerland}

\begin{abstract}
{Erosion shapes our landscape and occurs when a sufficient shear stress is exerted by a fluid on a sedimented  layer. 
What controls erosion at a microscopic level remains debated, especially near the threshold forcing where it stops.  
Here we study experimentally the collective dynamics of the moving particles, using  a set-up where the system spontaneously evolves toward the erosion onset. We find that 
the  spatial organization of the erosion flux is  heterogeneous in space, and  occurs along channels of local flux $\sigma$ whose  distribution  displays scaling near threshold and follows $P(\sigma)\sim J/\sigma$, where $J$ is the mean erosion flux. Channels are strongly correlated in the direction of forcing but not in the transverse direction. We show that these results quantitatively agree with a model where the dynamics is governed by the competition of disorder (which channels mobile particles) and particle interactions (which reduces channeling). These observations support that for laminar flows, erosion is a dynamical phase transition  which shares similarity with the plastic depinning transition occurring in dirty superconductors. The methodology we introduce here could be applied to probe these systems as well. 
}
\end{abstract}

\maketitle


{T}he response of erodible granular beds to shearing flows is of central importance in numerous natural phenomena such as sediment transport in rivers and estuaries, the evolution of mountains and landscapes, and the formation of dunes in the desert or underwater.  It  also affects many engineering processes such as slurry transport in mining or petroleum industries. 
However, and despite more than a century of studies, it  still lacks a complete fundamental understanding.
One of the essential issues is to describe the onset of solid flow. The incipient motion of the grains is controlled by  the Shields number, $\theta=\tau_b/(\rho_p-\rho_f) g d$, which is the shear stress $\tau_b$ induced by the fluid at the top of the bed scaled by the hydrostatic pressure-difference across the grains of diameter $d$. Here $\rho_p$ and $\rho_f$ are the density of the solid and the fluid, respectively, and $g$ the acceleration due to gravity.
One observes  a critical Shields number $\theta^c$ below which motion stops \cite{Buffington97}, following a  first transitory and intermittent regime in which the granular bed is continually reorganizing~\cite{Charru04}. This ageing or armoring of the bed leads to 
a saturated state of the bed   independent of its preparation \cite{Charru04,Loiseleux05,Ouriemi07,Derksen11,Kidanemariam14}. 
Once a stationary state is reached, the rate $J$ of particle transport above this threshold follows  $J\sim (\theta-\theta^c)^\beta$ with $\beta\in [1,2]$, as reviewed in \cite{Ouriemi09}. 

Several approaches have been introduced to describe these observations.   Bagnold~\cite{Bagnold66a} and followers \cite{Chiodi14}, emphasize the role of hydrodynamics. In their view,  moving particles carry a fraction of the total stress  proportional to their density $m$, such that the  bed of static particles effectively remains at the critical Shields number. The hydrodynamic effect of a moving particle on the static bed is treated on average, which neglects fluctuations. Erosion-deposition models \cite{Charru04} are another kind of  mean-field  description, which emphasize instead that moving particles can fill up holes in the static bed, leading to the armoring phenomenon described above. Deposition and erosion are modeled by rate  equations, which implicitly assumes that the moving particles visit the static bed surface entirely. More recently, collective effects have been emphasized. In \cite{Houssais15} it was proposed that the erosion threshold is  similar to the jamming transition that occurs when a bulk granular  material is sheared \cite{Houssais15}. Finally, two of us \cite{Yan16a} have proposed that the competing effects of  bed disorder and  interactions between  mobile particles controls the erosion onset. 


New observations are required to decide which theoretical framework is most appropriate to the erosion problem, and for which conditions.  In this letter, we study experimentally the collective dynamical effects of the mobile particles near threshold,  by measuring and averaging the trajectories of all the grains on the top of the bed.  Previously, a few studies have explored particle dynamics, but they have focused on isolated trajectories   \cite{Charru04}. Here instead we analyze for the first time the spatial organization of the erosion flux. We use a set-up where the Shield number slowly and spontaneously decreases as erosion occurs, as also occurs in  gravel rivers \cite{Parker07}.  This effect allows us to investigate precisely the approach to threshold.  Strikingly, we find that after averaging over time, the flux does  not become homogeneous in space. Instead,  fluctuations remain important and particles follow favored meandering paths. As the threshold is approached from higher Shields number, we find that most of the erosion flux is carried only by a few channels within the bed. Quantitatively, the distribution $P(\sigma)$ of local flux $\sigma$ in different channels is found to be extremely broad and to follow a power-law distribution $P(\sigma)\sim 1/\sigma$. Moreover, channels are  uncorrelated in the direction transverse to the flow, but display power-law correlations decaying as the inverse square root of the distance in the longitudinal direction. We perform a detailed comparison between these observations and the model introduced in  \cite{Yan16a}, and find quantitative agreements for a wide range of flows spanning from the viscous to the inertial regimes. Our work thus demonstrates the key role of disorder and particle interactions on the erosion threshold, and the need to use a framework that goes beyond mean-field approaches. In addition, it opens new ways to study dynamical phase transitions where both interactions and disorder are key, as is the case for the plastic depinning of vortices in dirty superconductors \cite{Kolton99,Watson96,Reichhardt16}  or skyrmions \cite{Reichhardt15}, in a setting where table-top experiments can be performed.  
%

\begin{figure}[htbp]
\begin{center}
\includegraphics[width=8.5cm]{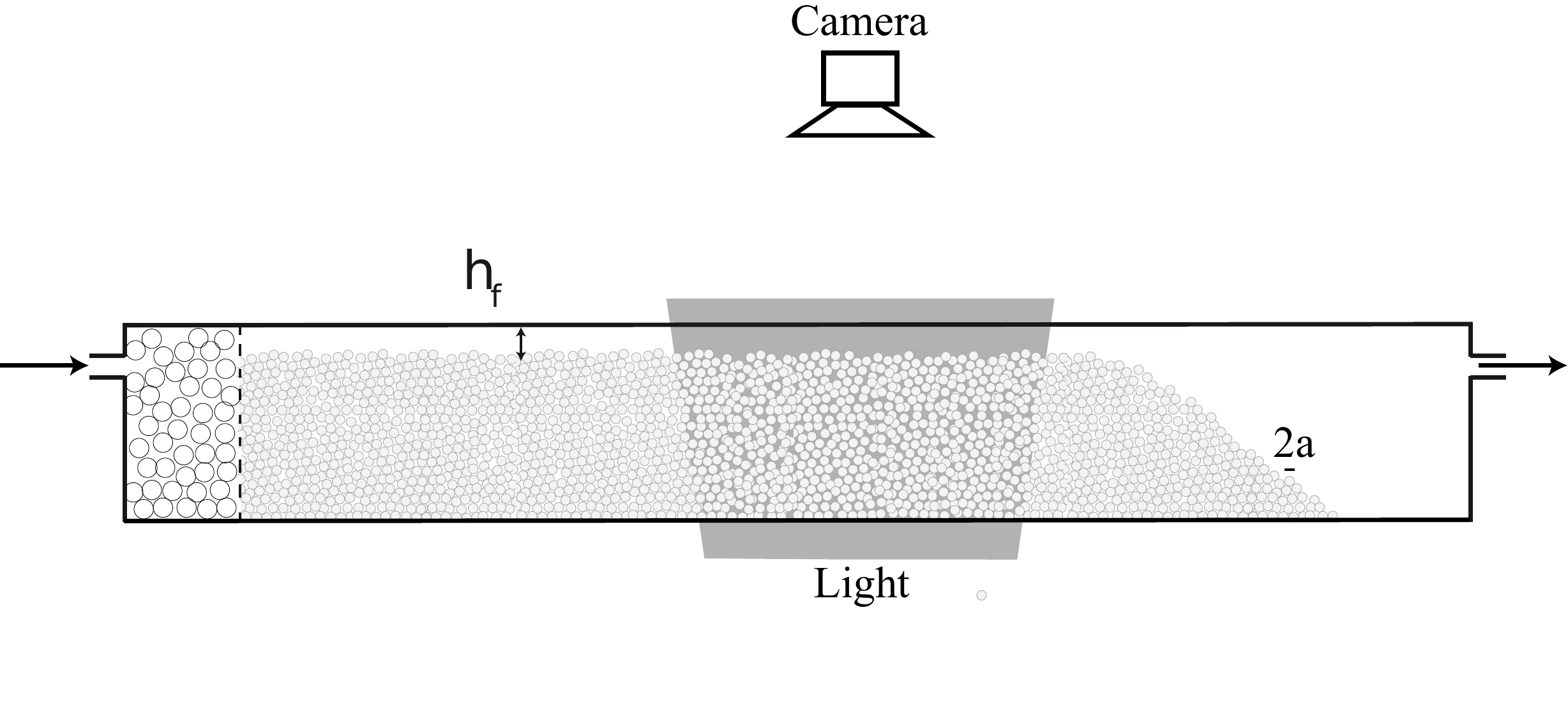}
\end{center}
\caption{Cross-section sketch of the experimental setup. The model flume apparatus consists of a rectangular perspex channel. It is filled with spherical particles of radius $a$ while leaving an empty buffer space in the downstream region. When a constant flow rate is imposed, eroded particles fall out into the empty buffer space. This leaves an upstream region exhibiting a flat fluid-particle interface, the height $h$ of which decreases with time until cessation of motion. A test section of this flat fluid-particle interface is imaged by a camera and the real-time positions and velocities of the moving particles are collected. }
\label{fig:setup}
\end{figure}

\subsection{Experimental setup}
In gravel rivers, erosion occurs until the fluid stress at the top of the river bed reaches its threshold value \cite{Parker07}. We use this effect and perform experiments in a model sediment river in which the Shield number continuously decreases as erosion occurs and eventually stops. 
In this set-up, the distance to threshold can be accurately monitored by measuring the particle flux $J$, which slowly decreases with time until it vanishes.

We use a model flume apparatus consisting of a rectangular perspex channel (height 3.5 cm, width 6.5 cm, and length 100 cm), see figure \ref{fig:setup}. We fill up the channel entrance with a granular bed of acrylic spherical particles (of radius $a=1$~mm and density $\rho_p=1.19$~g.cm$^{-3}$) while leaving an empty buffer space near the outlet. In order to cover both the viscous and inertial regimes of flows, this granular bed can be immersed in two different fluids, water (of viscosity $\eta=1.0$~cP and density $\rho_f= 1.00$~g.cm$^{-3}$) and a mixture of water and UCON oil (of viscosity $\eta= 44.5$~cP and density $\rho_f=1.06$~g.cm$^{-3}$). A given flow rate driven by a gear pump is then imposed and kept constant for the duration of each experimental run.  At the inlet of the channel, the fluid flows through a packed bed of large spheres, providing a homogeneous and laminar flow. At the outlet, the fluid is run into a thermostated fluid reservoir, which ensures a constant temperature of 25~$^{\circ}$C across the whole flow loop.

In this geometry,  eroded particles fall out into the empty buffer space at the outlet. This leaves an upstream region exhibiting a flat fluid-particle interface, whose height  decreases with time until cessation of motion. At constant fluid flow, $\theta$ decreases with the thickness of the fluid layer  $h_f$ (which increases with time) until the threshold of motion is reached from above  \cite{Ouriemi07}. The experimental measurements are undertaken in the vicinity of the onset of motion, i.e. in a flow regime where only the particles located in the top one-particle-diameter layer of the bed are in motion.  They consist of recording sequences of images of the top of the bed in a test section of the channel using a specially-designed particle-tracking system, see details in {\it methods}. The real-time positions and velocities of the moving particles are collected and both local and total particles fluxes, $\sigma$ and $J$ respectively, are inferred as will be described below. 


\begin{figure*}
\begin{center}
\includegraphics[width=7cm]{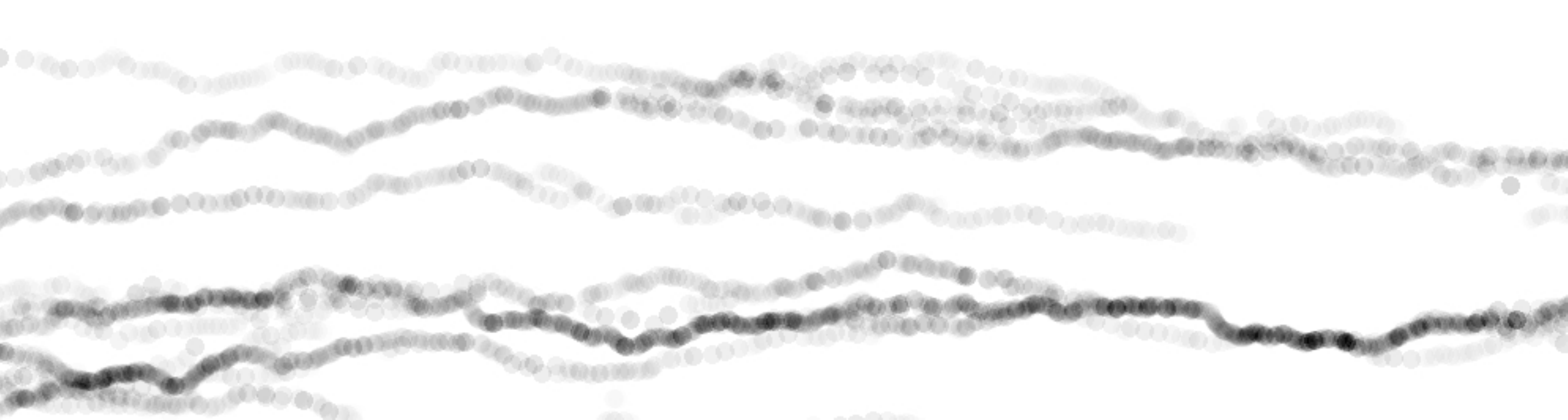}
\includegraphics[width=7cm]{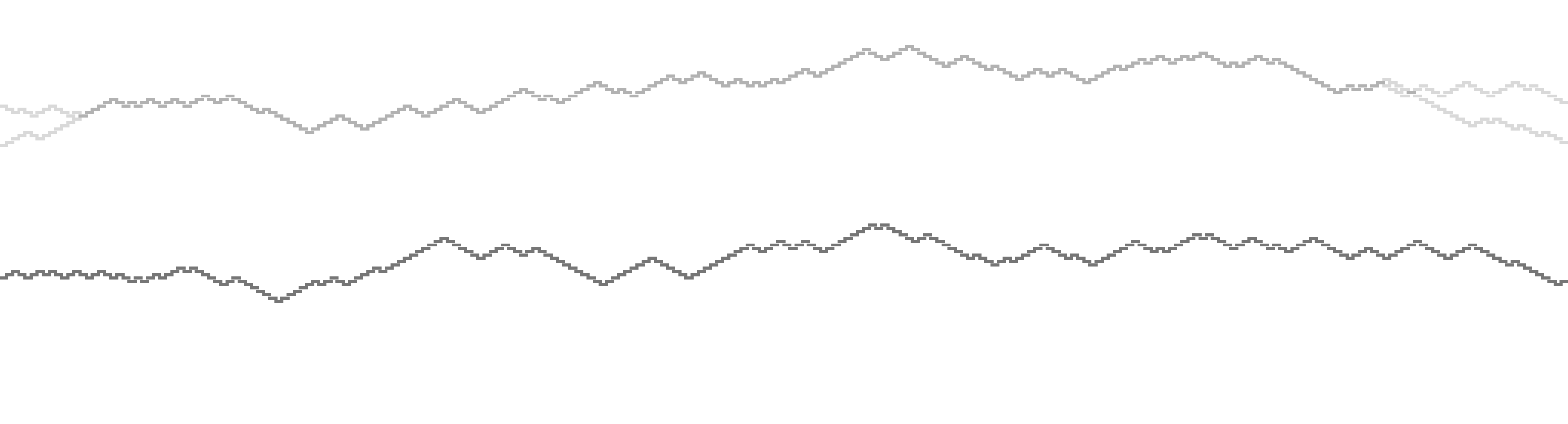}\\
\includegraphics[width=7cm]{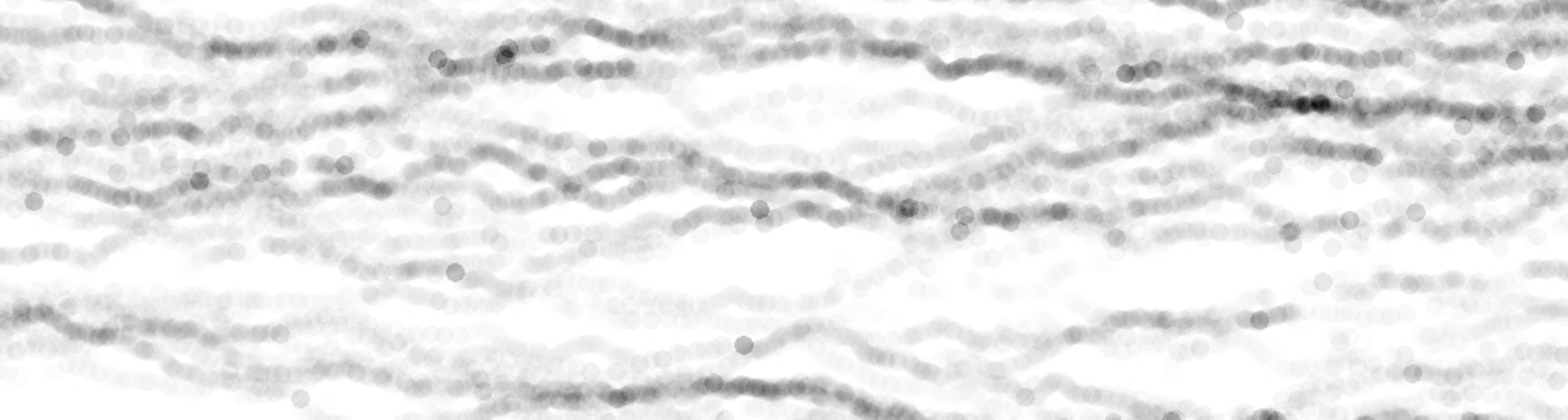}
\includegraphics[width=7cm]{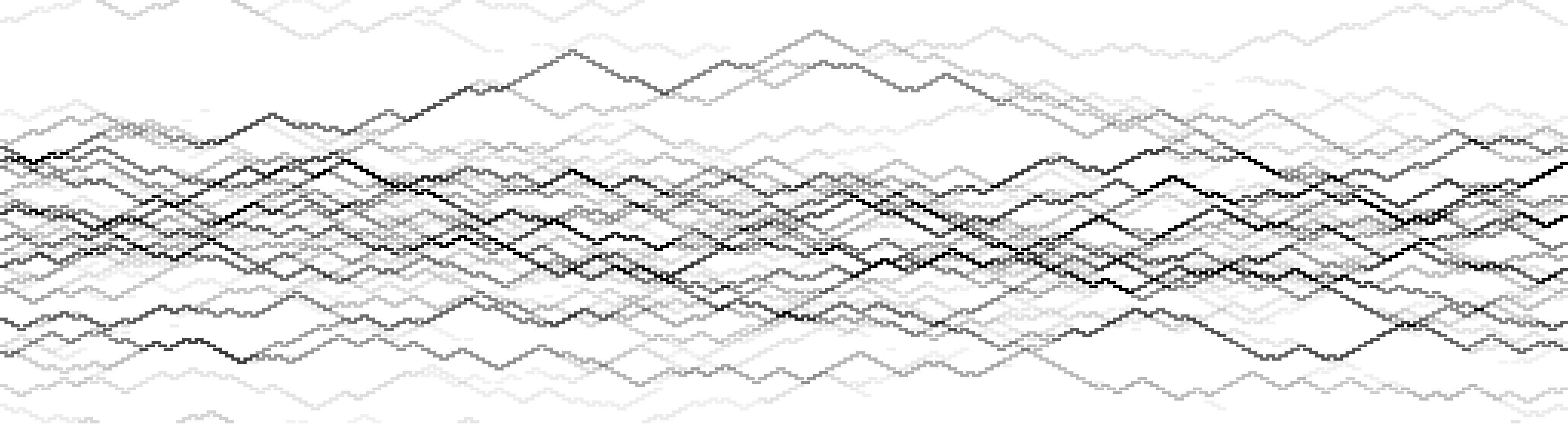}\\
\includegraphics[width=7cm]{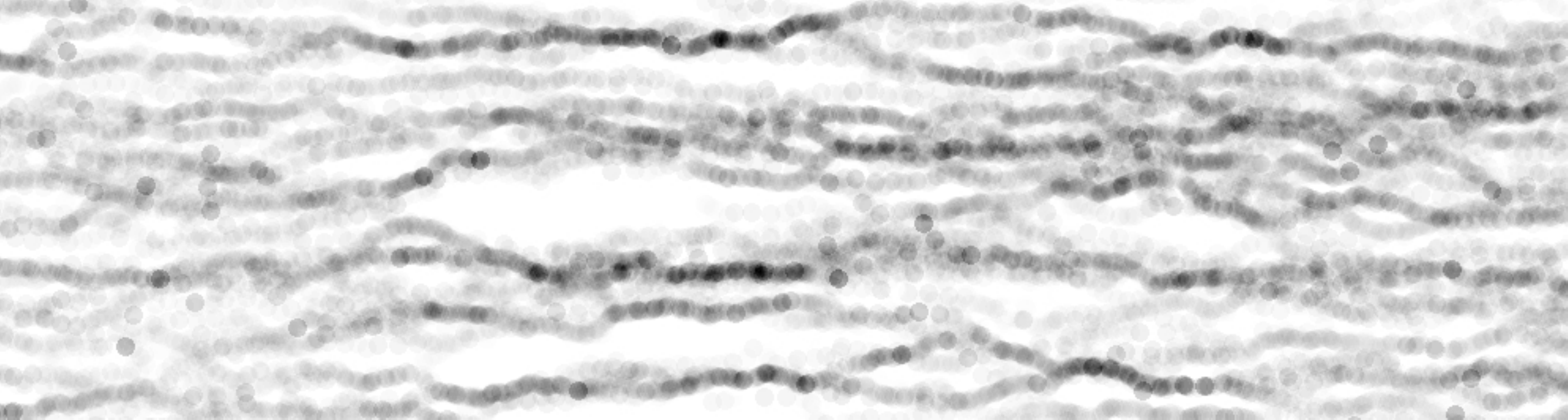}
\includegraphics[width=7cm]{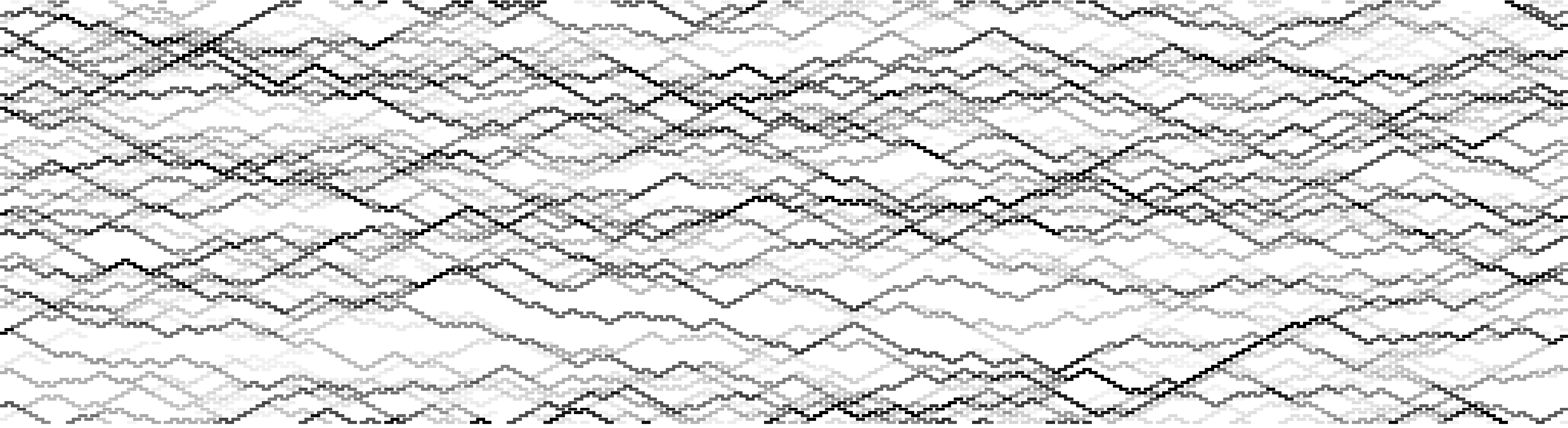}\\
\includegraphics[width=7cm]{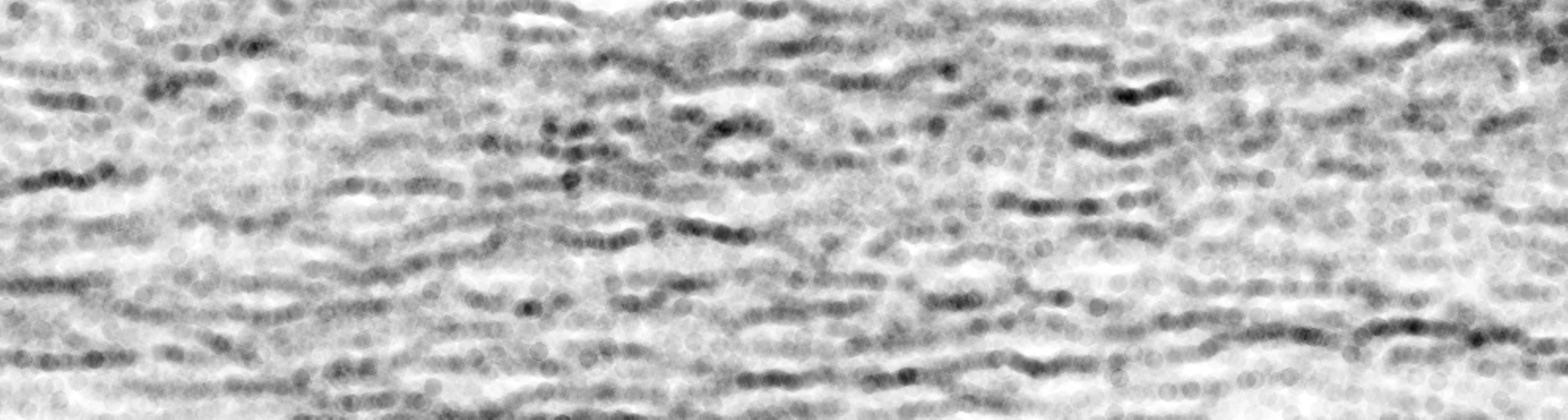}
\includegraphics[width=7cm]{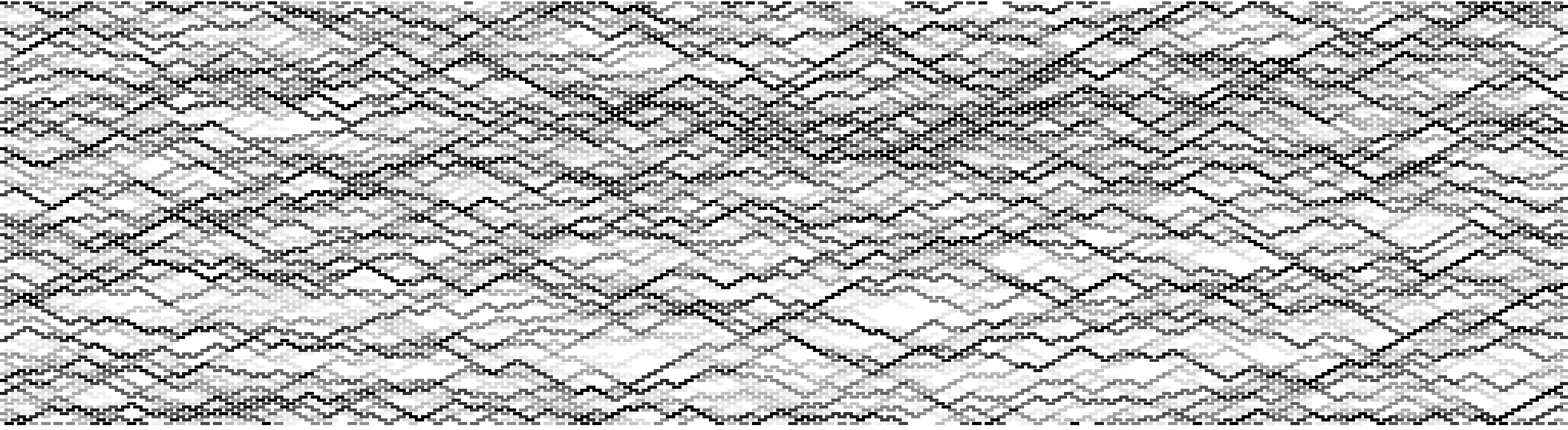}
\end{center}
\caption{Typical channeling patterns: experiments (left) at $J/J_{\rm max} = 0.09, 0.35, 0.54,$ and $0.95$ (from top to bottom), where $J_{\rm max}=0.081$ and model (right) at $J/J_{\rm max}=0.18,0.25,0.46,$ and $0.78$ (from top to bottom), where $J_{\rm max}=0.25$. Left: Experimental trajectories of the moving particles where the grayness indicates the magnitude of the local flux (see also online movies in S.I.  for both the pure water and the water-UCON mixture at different $J/J_{\rm max}$). Darker paths corresponds to paths which are more often visited by particles.  Right: The solid lines show the local fluxes $\sigma$ along the edges, whose magnitudes measured in the steady state are indicated by the grayscale of the lines.}
\label{fig:channelling}
\end{figure*}

{\bf Channelling pattern:}
Using particle tracking, the downstream and lateral velocities of each moving particle, $u$ and $v$ respectively, are obtained. Time-averaging over all the moving particles in the $N$ processed images is then performed. The mean transverse velocity is found to be zero while the mean downstream (or longitudinal) velocity is approximately constant for all the runs for a given fluid. This is consistent with earlier findings that as threshold is approached, the density $m$ of moving particles vanishes, but not their average speed \cite{Charru04,Lajeunesse10,Duran14}.  Averaging over all runs yields a mean longitudinal velocity $U= 2.2$~mm/s for the water-Ucon mixture and $=36.1$~mm/s for water. 
\footnote{This value of $U$ can be simply recovered by balancing the drag force $C_D \rho_f \pi a^2 U^2/2$ on a particle with the friction force on the top of the bed $4 \mu \pi a^3 (\rho_p-\rho_f) g/3$, where $C_D=[24 / Re_p][1+0.15  Re_p^{0.687}]$ is the Schiller-Naumann correlation for the drag coefficient with the particle Reynolds number defined as $Re_p= \rho_f aU/\eta$ and $\mu \approx 0.33$ is the friction coefficient, the value of which is in agreement with that found in previous work for suspensions \cite{Cassar05,Boyer11}. The particle Reynolds number is $Re_p=0.05$ for the water-Ucon mixture and $Re_p=36.10$ for pure water.}

From the measurement of these local particle velocities, the normalized local particle flux:
\begin{equation}
\sigma (i,j) = \frac{1}{N} \sum_{\rm particles} (u/U),
\label{eq:sigma}
\end{equation}
can be inferred at a given site $(i,j)$ within a box having the size of one pixel in the image (one pixel is $\approx 0.15$~mm). Note that the sum of the normalized local velocities $u/U$ is undertaken over all the moving particles in the $N$ processed images.

An example of the flux spatial organization $\sigma (i,j)$ is given in figure \ref{fig:channelling}~(left) as a grey scale level. One of our central findings is that after time averaging, the erosion flux is not uniform.  Darker regions indicate paths which are more often visited by particles.
 Close to incipient motion, only a few channels are explored by the particles, see the top image of figure \ref{fig:channelling}~(left). Further from threshold, a greater number of preferential paths are followed and eventually  the particle trajectories cover the whole bed surface, see the bottom images of figure \ref{fig:channelling}~(left).

\begin{figure*}
\begin{center}
\includegraphics[width=7cm,height=5.3cm]{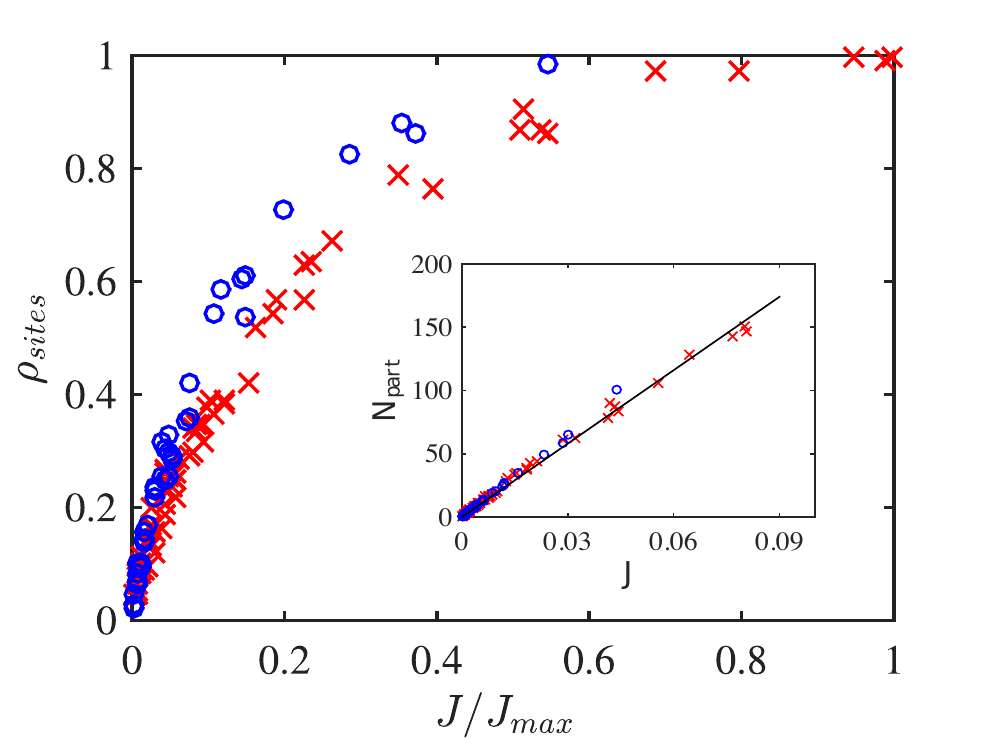}
\includegraphics[width=7cm,height=5.3cm]{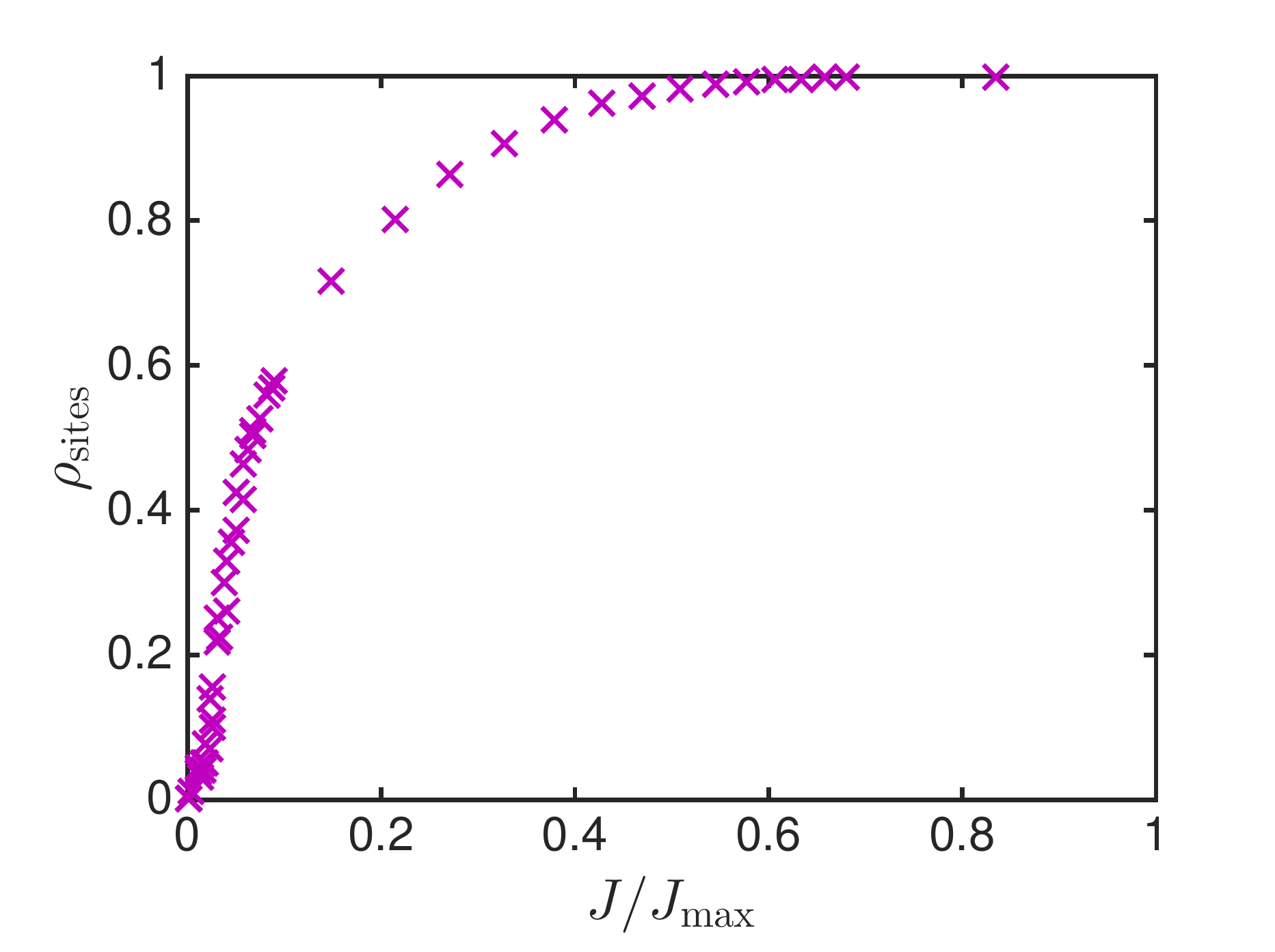}
\end{center}
\caption{Density of visited sites $\rho_{\rm sites}$ versus particle flux $J$ scaled by the maximum value $J_{\rm max}$ : experiments (left) using the water-UCON mixture ($\times$) and pure water ($\circ$), and model (right). The inset of the left graph shows that number of moving particles is linear in particle flux.}
\label{fig:densityrhos}
\end{figure*}

{\bf Surface visited by moving particles:}
We now quantify how the mean number of visited sites depends on the distance to the erosion threshold. The total normalized particle flux $J$ is  defined as the spatial average of the local particle flux $\sigma$ over the $N_{\rm pixels}$ boxes in the image:
\begin{equation}
J =  \frac{1}{N_{\rm pixels}} \sum_{i,j} \, \sigma(i,j).
\label{eq:J}
\end{equation}
Since the Shields number cannot be directly measured when the bed is viewed from above, the total flux $J$, which is a continuous function of the Shields number $\theta$, is chosen as the control parameter of the experiment. As $J$ increases, we find that the number of sites explored by the particles increases and eventually saturates when the whole surface of the test section is visited for a value $J_{\rm max}=0.081$. In Fig.\ref{fig:densityrhos}~(left)  the surface density of visited sites, $\rho_{\rm sites}$ (defined as the fraction of visited pixels in images such as those shown in Fig.\ref{fig:channelling}) is plotted versus $J$. Interestingly, the data recorded in the viscous ($\times$) and inertial ($\circ$)  regimes (i.e. data obtained with the water-Ucon mixture and pure water, respectively) have the same trend and even are close to collapsing onto the same curve. The inset of figure \ref{fig:densityrhos}~(left) shows that the number of moving particles is linear in particle flux and vanishes at threshold  both for the viscous and inertial data.


\begin{figure*}
\begin{center}
\includegraphics[width=7cm,height=5cm]{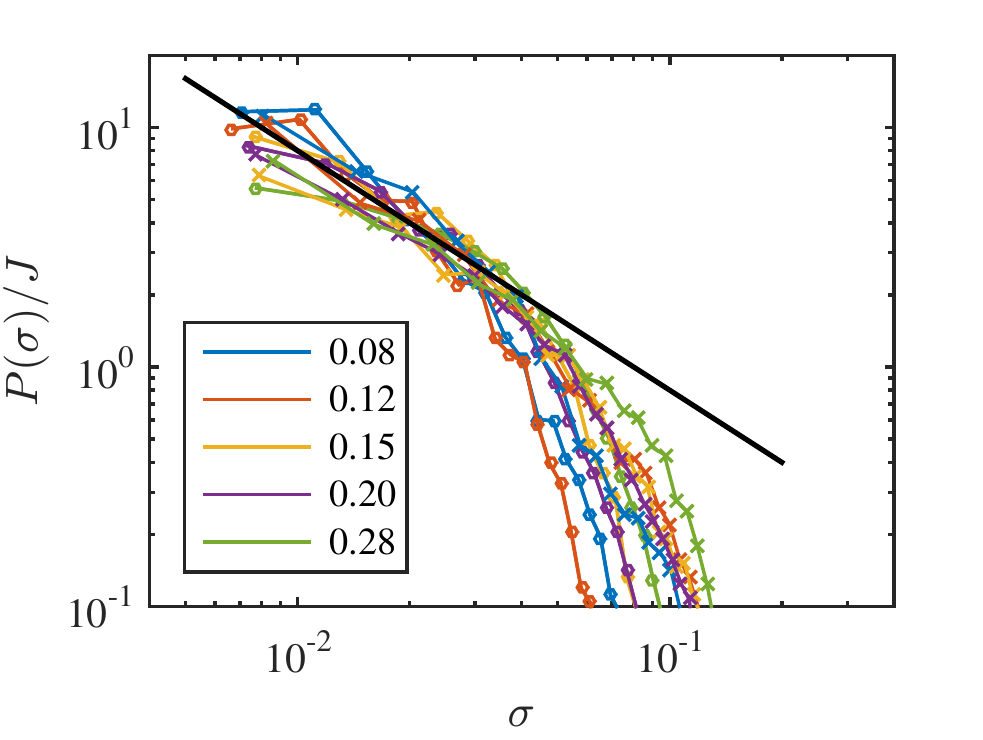}
\includegraphics[width=7cm,height=5cm]{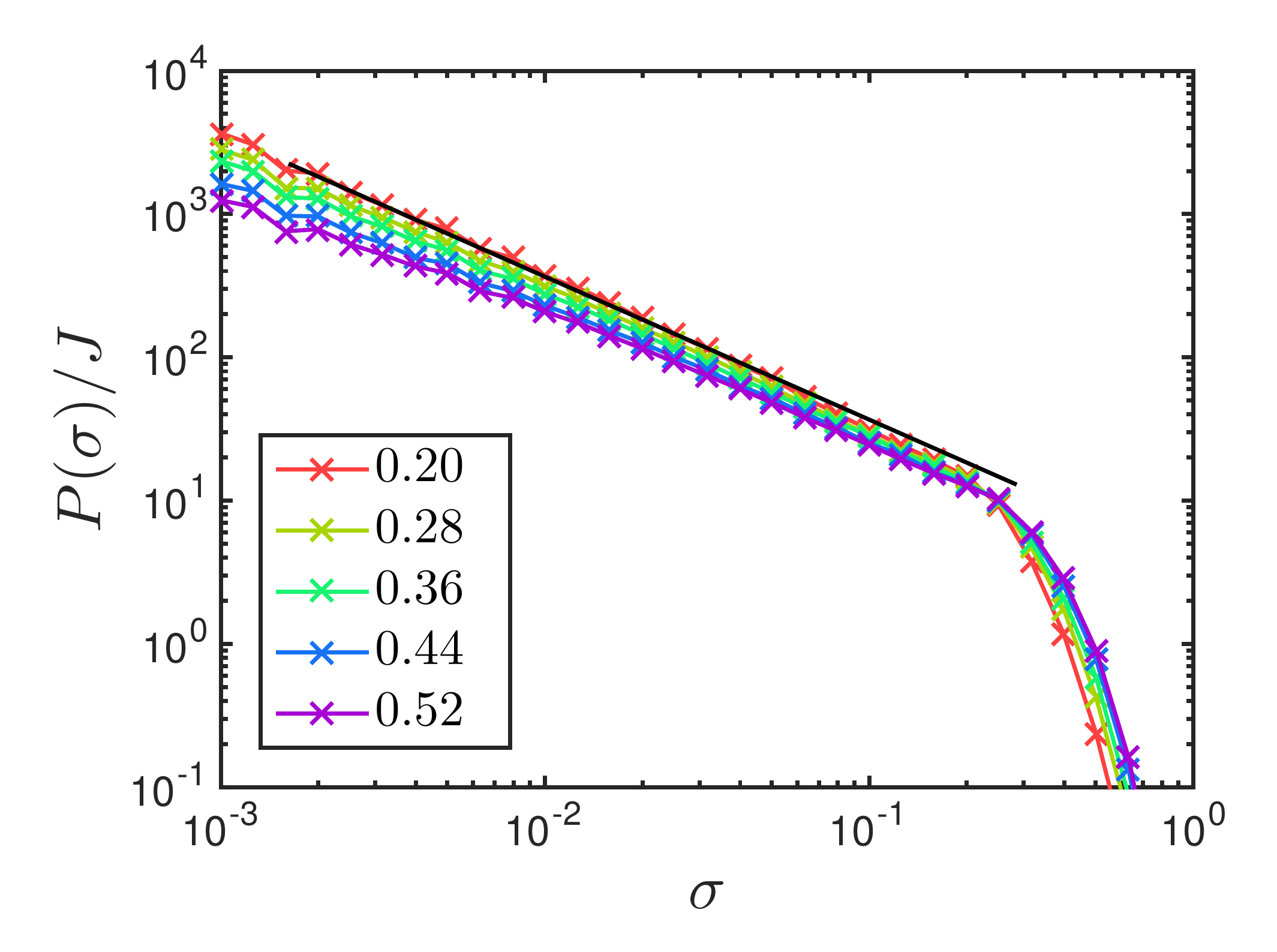}\\
\includegraphics[width=7cm,height=5cm]{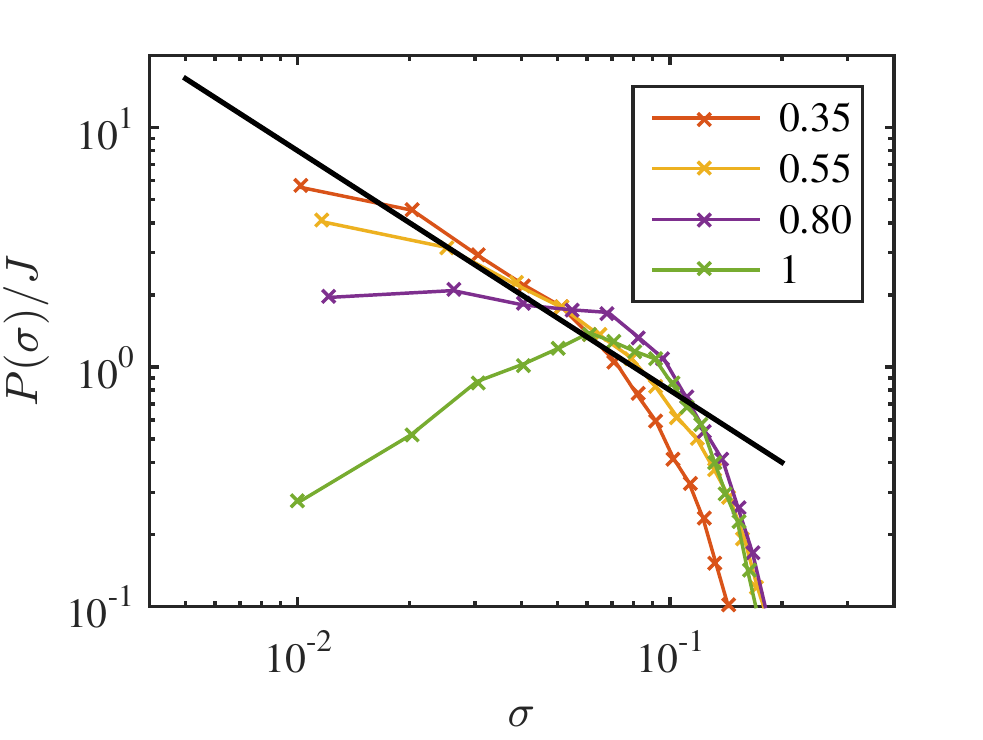}
\includegraphics[width=7cm,height=5cm]{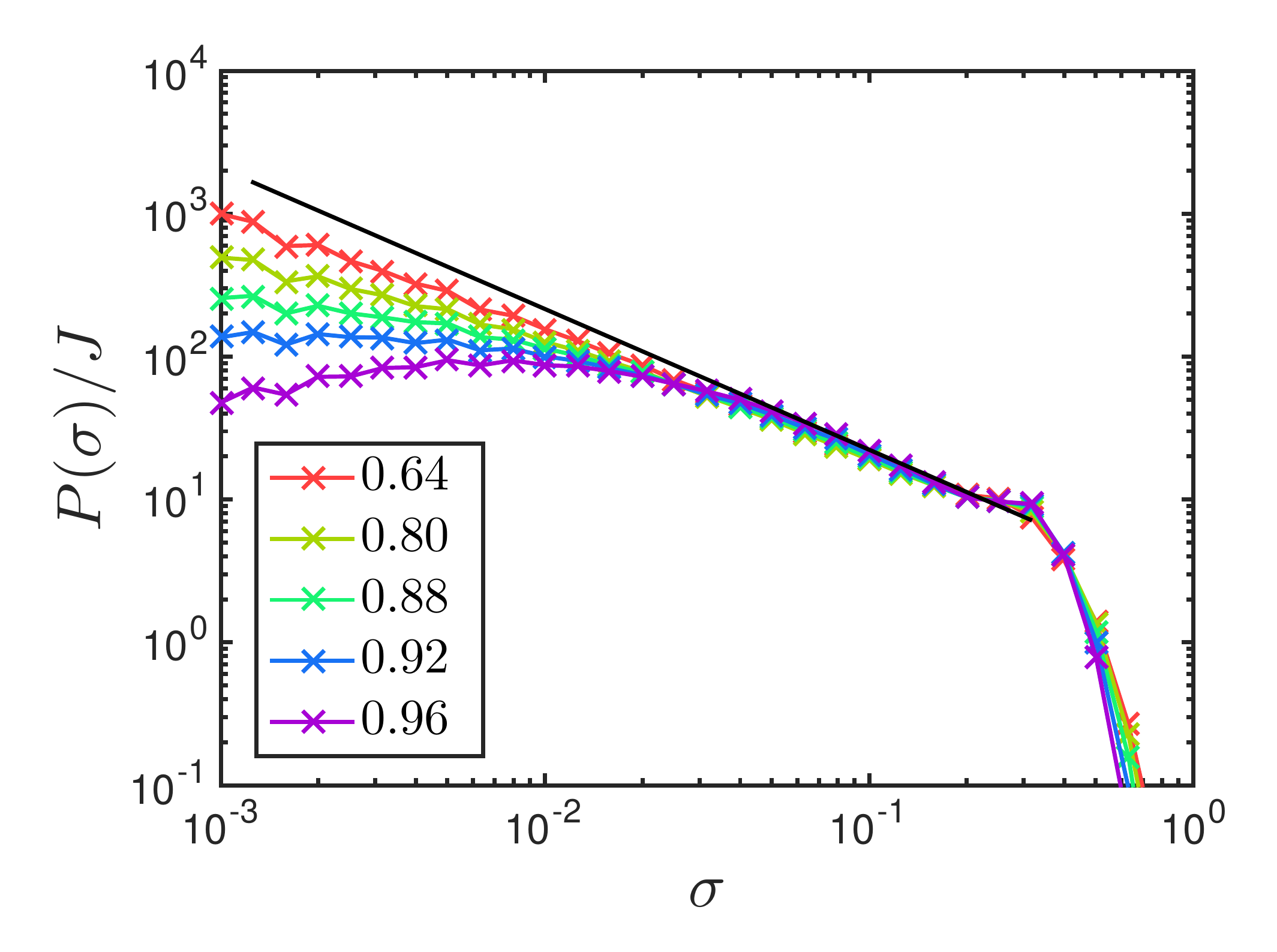}
\end{center}
\caption{Probability density of local fluxes $P(\sigma)/J$ for small flux $J$ (top) and large $J$ (bottom): experiments (left) using the water-UCON mixture ($\times$) and pure water ($\circ$), the theoretical model (right). The values in the legend are $J/J_{\rm max}$, and the black solid lines scale as $\sigma^{-1}$.}
\label{fig:distributionPsigma}
\end{figure*}

{\bf Distribution of channel strengths:}
To quantify the spatial organization of the erosion flux, we compute the distribution $P(\sigma)$ of the local particle fluxes $\sigma$ as $J$ is varied. Our key findings are shown in Figure \ref{fig:distributionPsigma}~(left): 

(i)  close to threshold, the different curves $P(\sigma, J)$ can be collapsed  using the functional form $P(\sigma, J)=J f(\sigma)$.  Such scaling collapse is reminiscent of a continuous critical point. Note that this collapse holds in the range of local flux $\sigma$ probed experimentally, but it cannot hold always, since the distribution must integrate to one, as discussed below.  

(ii)  For both the viscous ($\times$) and inertial ($\circ$) regimes, the function $f(x)$ is  well fitted by the function $1/x$ (solid lines in the graphs), leading to:
\begin{equation}
P(\sigma)/J \propto \sigma^{-1},
\label{eq:fitPsigma}
\end{equation}
Such a broad distribution is  characteristic of a channeling phenomenon, for which some sites are almost never visited, while others are visited very often. Eq.\ref{eq:fitPsigma} has no scale,  indicating that the channel pattern is a fractal object.  Obviously,  at large $\sigma$ this distribution is cut-off,   as shown in the top graph of figure \ref{fig:distributionPsigma}~(left). This simply indicates that there is a maximum possible flux a site can carry, if particles have a finite speed. More surprisingly, Eq.(\ref{eq:fitPsigma}) together with the constraint that $P(\sigma)$ integrates to one indicates the presence of a cut-off $\sigma_{min}\sim e^{-1/J}$, a quantity so small however that it does not appear in our observations at small $J$.
However for $J/J_{\rm max}\gtrsim 0.037$, the scaling form of equation (\ref{eq:fitPsigma}) breaks down at small $\sigma$, see bottom graph of figure \ref{fig:distributionPsigma}~(left). 

{\bf Spatial correlations of the channel network:}
We now turn to the analysis of the spatial correlations of the particle flux, defined as:
\begin{eqnarray}
C_T(\Delta j) & = & \langle \sigma(i,j)\sigma(i,j+\Delta j)\rangle_c / \langle \sigma(i,j)\sigma(i,j)\rangle_c,\\
C_L(\Delta i) & = & \langle \sigma(i,j)\sigma(i+\Delta i,j)\rangle_c / \langle \sigma(i,j)\sigma(i,j)\rangle_c,
\label{eq:CT}
\end{eqnarray}
in the transverse and longitudinal directions, respectively.  Here the symbol $\langle\bullet \rangle$ indicates a spatial average, and $\langle xy\rangle_c\equiv \langle xy\rangle- \langle x\rangle\langle y\rangle$. 

Figure \ref{fig:CT}~(left) shows that
there is no correlation in the transverse direction beyond $\approx 2a$ whereas long-range correlations appear in the longitudinal direction beyond $\approx 2a$ (delimited by a dashed line in the graphs). For small $J$, the decay can be well represented by a power-law, $(\Delta i/a)^
\alpha$ with $\alpha\approx-0.5$ (solid line in the bottom graph), and is independent of $J$. This observation further supports  that the channel pattern is fractal with no characteristic length scales.  For larger $J$, the decay deviates from this law and becomes stronger with increasing $J$.


\subsection{Theoretical model}
  We now show that these observations  {\it quantitatively}  agree with a theory incorporating two  ingredients: (i) the channelling induced by the disorder (resulting from the presence of an essentially static bed) and (ii) the interaction among mobile particles. Why the first ingredient implies the second can be argued as follows: the trajectory of a single mobile particle must overall follow the main direction of forcing, but will meander because it evolves on a bed  which is disordered and  static. Thus there are favored paths  which particles follow. If there are several mobile particles, this effect of the disorder tends to channel particles together along these paths. If mobile particles were not interacting, nothing would stop this coarsening to continue, and eventually  all particles would be attracted to the same optimal path. Obviously,   this scenario is impossible for a large system as the density along the favored path would be much larger than unity. Particle interaction is thus key to limit channeling.  Interactions result in two effects: first, a mobile particle cannot move into a site already occupied by another particle. Second,  another particle can push on a mobile one and can deviate the latter from its favored path.

In \cite{Yan16a}  these effects were incorporated in  a model where both space and time were discretized, and where inertial effects as well as long-range hydrodynamic interactions were neglected. The static bed is treated as a frozen background or random heights $h_l$, where $l$ labels the different sites of a square lattice.
A fraction $n$ of the lattice sites are occupied by particles that can move under   conditions discussed below. 
The direction of forcing is along the lattice diagonal, indicated by the arrow in Fig.~\ref{fig:model}. There are two inlet bonds and two outlet bonds for each node. Bonds $l\to m$ are directed in the forcing direction, and characterized by an declination $\theta_{l\to m} = h_l-h_m$. For an isolated particle on site $l$,  motion occurs if there is an outlet for which $\theta+\theta_{l\to m}>0$, where  $\theta$ is the magnitude of the forcing  acting on all  particles. Flow occurs along the steepest of the two outlets, resulting in channeling. When particles move, they do so with a constant velocity, thus the flux $J$ is simply the density of mobile particles, and is bounded by $J_{\rm max}=n$ (the value of $n$ does not affects the critical properties for $J\ll1$, in our figures $n=0.25$). 

Finally, particles  cannot overlap, but they can exert  repulsive forces on particles below them. Such forces can un-trap a  particle that was blocked, but can also deviate a moving particle from its course, as illustrated in Fig.~\ref{fig:model}. The path of a particle thus  depends also on the presence of particles above it.  As long as these features are present, we expect the model predictions to be  independent of the details of the interactions.  The detailed implementation of forces are presented in {\it methods}. 

{\bf Numerical results:}
As shown in Figs.\ref{fig:channelling}, the channelling map generated by the model reproduces qualitatively the experimental ones. 
Likewise, the dependence of the surface visited by mobile particles on the flux $J$ shown in Fig. \ref{fig:densityrhos} closely 
matches experimental finding. 

Our central result however is that this agreement is {\it quantitative}. As shown in \ref{fig:distributionPsigma}, both the model and the experiment obtain the same  form for $P(\sigma)\sim J/\sigma $. This result is unusual. It  is not captured for example by simple  models of river networks \cite{Dhar06} which also display some channeling.  We are not aware of any alternative theory making such a prediction. As $J$ increases, scaling breaks down and  $P(\sigma)$ becomes peaked both in experiments and in the model.

\begin{figure*}
\begin{center}
\includegraphics[width=7cm,height=5cm]{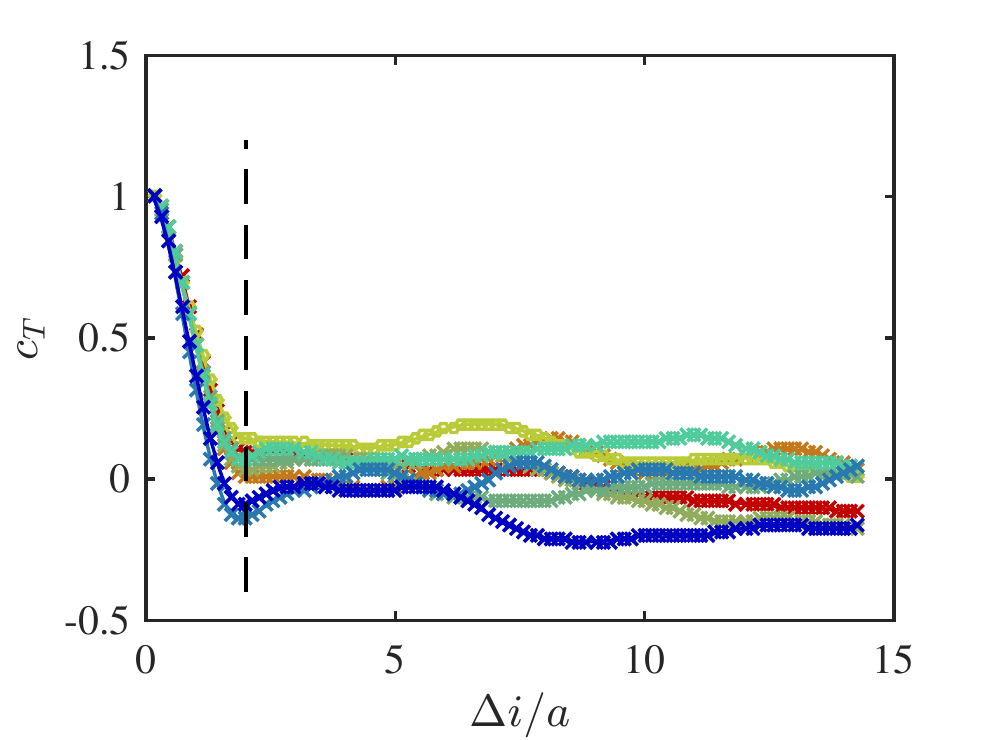}
\includegraphics[width=7cm,height=5cm]{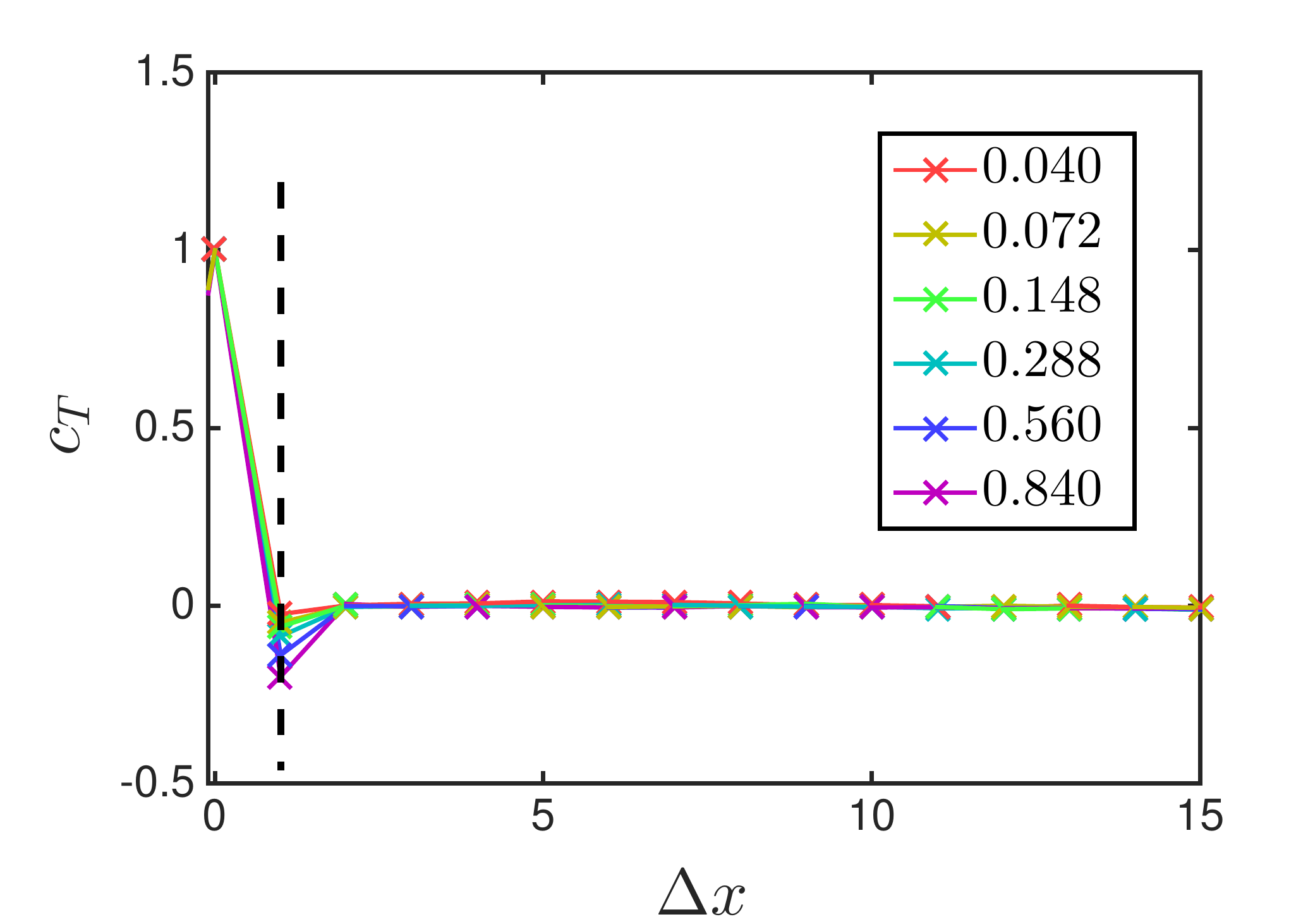}\\
\includegraphics[width=7cm,height=5cm]{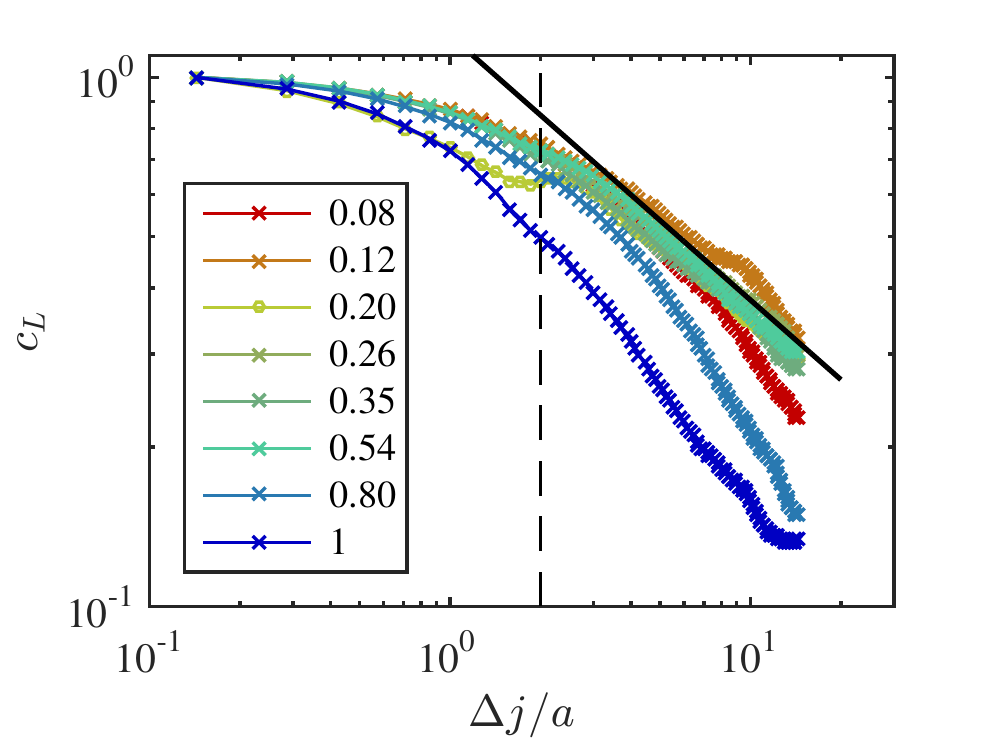}
\includegraphics[width=7cm,height=5cm]{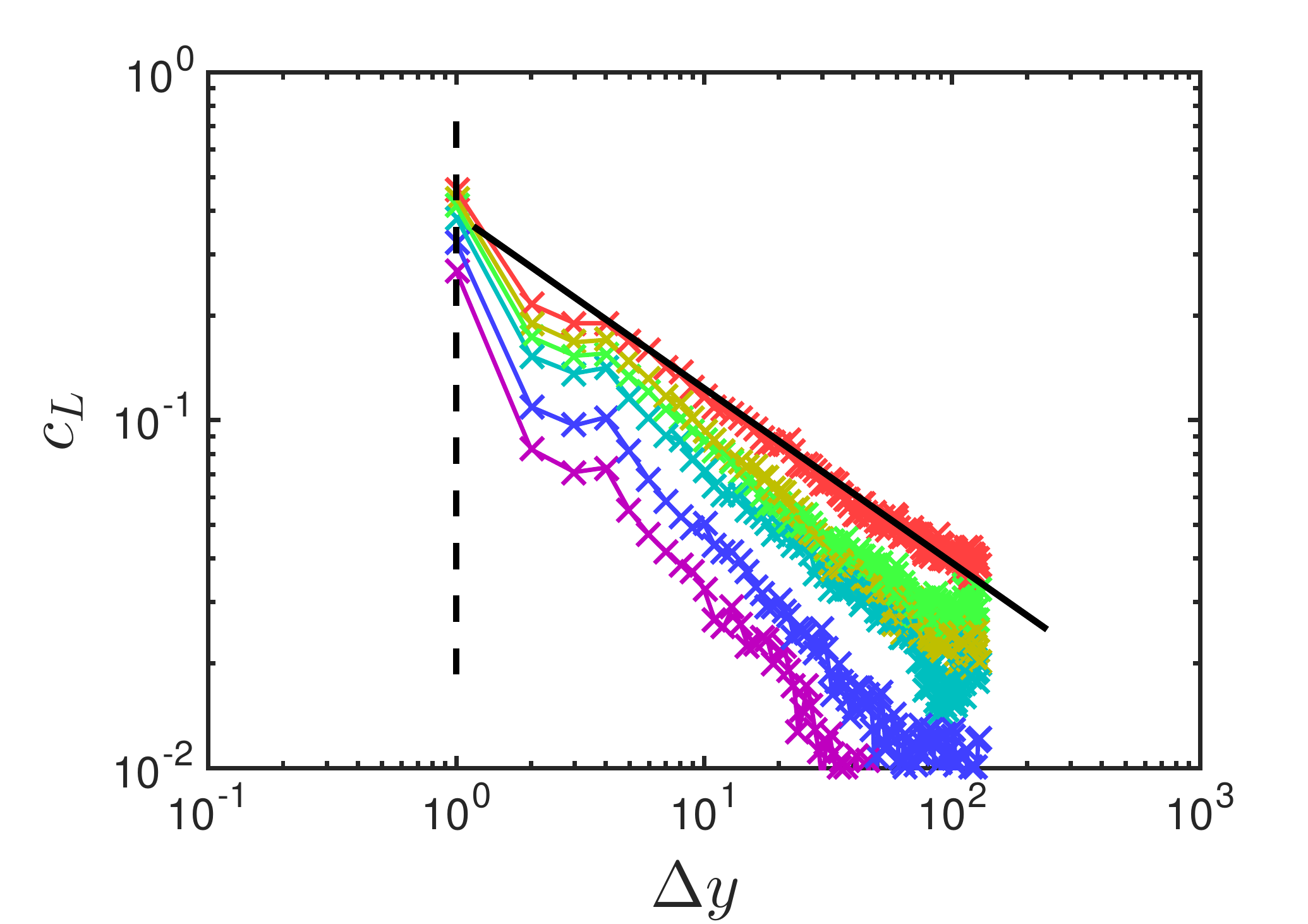}
\end{center}
\caption{Transverse $C_T(\Delta j)$ (top) and longitudinal correlations $C_L(\Delta i)$ (bottom) for various $J$: experiments including both data for water-UCON mixture and water (left)   and model (right).  The black dashed lines indicate the distance $2a$ in the experiment and one lattice constant in the model. The black solid lines show $(\Delta j/a)^{-0.5}$ in the experiment and $\Delta y^{-0.5}$ in the model where the space is measured in the unit of the lattice constant.}
\label{fig:CT}
\end{figure*}

The same quantitative agreement is found for the spatial correlations of the channel strength  $C_T$ and $C_L$, as shown in Figs.~ \ref{fig:CT}: there are essential no correlations  in the transverse direction, but correlations are  long-range and decay as $1/\sqrt{\Delta y}$ at small $J$, where $\Delta y$ is the distance between two sites along the flow direction. 

%

\subsection{Conclusion}
We have shown experimentally that near the erosion onset, the flow of particles is heterogeneous,
and concentrates into channels whose amplitude is power-law distributed. Such channels display long-range correlations in the main direction of flow, but no correlations in the transverse direction. 

These observations  are in striking agreement with a model where the particle dynamics is controlled both by the disorder of the bed of static particles, as well as local interactions between mobile particles.  This quantitative agreement supports that effects ignored in the model are irrelevant near the transition, at least for the regime of flow reported here. This includes  long-range hydrodynamic interactions, as well as the very slow creep flow of the granular bed below the mobile particles \cite{Houssais15}.

In the framework that emerges from our work, the interplay between disorder and interactions leads to  a dynamical phase separating between an arrested phase and a flowing one. The transition is continuous, as supported by the scale free channel organization near threshold reported here. Generally, near such transitions the dynamics is expected to be singular, and indeed the model predicts $J\sim (\theta-\theta_c)^\beta$ with $\beta=1$ \cite{Yan16a}. This exponent is consistent with previously experimentally reported values, but precise data measuring accurately $\beta$ would be very valuable to test this theory further. 

Finally, the proposed framework supports a direct comparison between the erosion threshold and other dynamical systems where interacting particles are driven in a   disordered environment \cite{Kolton99,Watson96,Reichhardt02, Pertsinidis08}. A classical example are type II superconductors in which the disorder is strong enough to destroy the crystallinity of the vortex lattice \cite{Kolton99,Watson96,Reichhardt16}. If the forcing (induced by applying a magnetic field) is larger than some threshold, vortices flow along certain favored paths, reminiscent of the dynamics reported here   \cite{Kolton99}, a phenomenon referred to as {\it plastic depinning} which is not well understood theoretically \cite{Reichhardt16}.  Previous theoretical  models of this phenomenon \cite{Watson96} did not consider that the interaction between particles can deviate them from their favored path. Such models lead to  channels whose amplitude  $\sigma$ is zero or one (i.e. $P(\sigma)$ is the sum of two delta functions), at odds with the broad distribution reported here. It would be very interesting to check if our framework applies to plastic depinning in general, by testing as we have done here if the distribution of channel strength is  indeed power-law, or bimodal. 

\subsection{Methods}

{\bf Experiment:}
The experimental measurements are performed in a channel test section of length $L= 150$~mm and width $W= 40$~mm, located at a distance of $\approx 500$~mm from the channel entrance. This test section is illuminated from below by an homogeneous light and imaged from above by a digital camera (Basler Scout) with a resolution of $1392 \times 1040$ pixels, see figure \ref{fig:setup}. For a given run, typically 3 to 4 sequences of typically 300 images are recorded. Note that the different sequences correspond to different decreasing bed height and thus to different decreasing particle flux $J$ until cessation of motion is reached. The images are recorded at a rate of 30 frames per second for water and at a rate of 3.75 frames per second for the water-Ucon mixture. The number of images $N$ which is eventually processed is chosen as to correspond to a travelled length of $193$~mm, i.e. $N=160$ for water and $N=257$ for the water-Ucon mixture. These images are then processed to infer real-time positions and velocities of the moving particles. First, for each image of a given sequence, the moving median grey-level image is calculated over a subset of 11 images surrounding the given image (the 5 preceding and following images in addition to the given image). This moving median image is then subtracted from the given image. This provides a new image which only highlights the moving particles.  Second, a convolution of this new image with a disk having the size of the particles is performed. The resulting maximum intensities yield the centers of the moving particles. Particle trajectories and velocities are finally calculated by using a simple particle-tracking algorithm which relied on the small displacement of the tracked particles between two sequential images by imposing an upper bound condition on particle displacement. Note that these conditions depend on the direction, i.e. the downstream and lateral bounds are smaller than the upstream bound. 

\begin{figure}
\centering
\includegraphics[width=7cm]{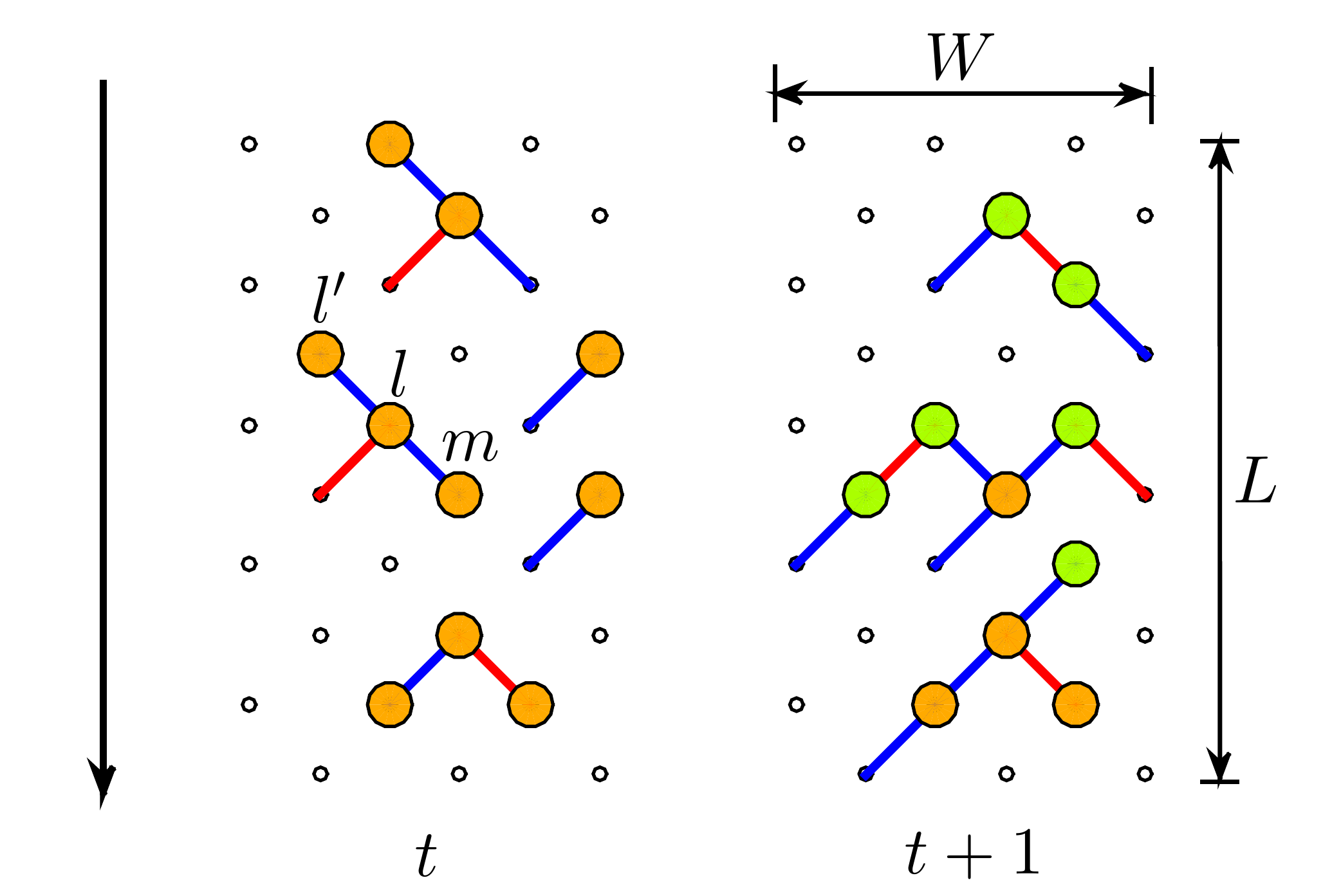}
\caption{Illustration of the model, embedded on a square lattice of length $L$ and width $W$. At each moment, each lattice site indicated by a small circle can accommodate at most one particle represented by a disc. The black arrow along the square diagonal indicates the downstream direction. Solid lines extended from the particles infer the outlets with positive forces. The outlet of the larger force is colored in blue, the smallest is in red. The green discs show the particles moved in the step $t$ (left) to $t+1$ (right).}
\label{fig:model}
\end{figure}
{\bf Model:}
Particles interact when they are adjacent. We denote by $f$ the unbalanced force acting on one particle, coming both from particles above it (if they are present), as well as from a combination of gravity and forcing. The force vector is decomposed into two scalar components along the two outlets: 
the component $f_{l\to m}$ on bond $l\to m$ is determined by:
\begin{equation}
\label{eq:force}
f_{l\to m} = \max(f_{l'\to l}+\theta_{l\to m}+\theta,0)
\end{equation}
where  $f_{l'\to l}$ is the unbalanced force on particle $l'$ in the direction of the bond $l'\to l$, in the same direction as $l\to m$, as depicted in Fig.~\ref{fig:model}.  If the site $l'$ is empty,  $f_{l'\to l}=0$.
That term $f_{l'\to l}$ captures  that if a particle pushes on another one below, the latter has a stronger unbalanced force in that direction. The term $\theta_{l\to m}+\theta$ characterizes the strength of the forcing with respect to the inclination of the link $l\to m$.



From a given state at time $t$, we first compute all the forces, illustrated by the red and blue lines in Fig.~\ref{fig:model}. Then particles which present non-zero unbalanced forces will move in the direction where the force is greatest, if that site below is empty. In practice, we start from the bottom row. For each row, the particles are moved to the unoccupied sites in the row below, starting from the largest unbalanced forces $f_{l\to m}$. Rows are updated one by one toward the top of the system. 
In our model we use periodic boundaries. After all $L$ rows (each of width $W$) have all been updated, time $t$ increases to $t+1$.

For a given Shields number $\theta$, we initialize the system with particles randomly positioned and study dynamic quantities in the steady state $t\to\infty$. In practice, we average the properties in $t\in[LW,2LW]$.
Our results are shown for $L=256$ and $W=64$.

\begin{acknowledgments}
We thank B. Andreotti,  D. Bartolo, P. Claudin, E. DeGiuli and J. Lin  for discussions.  M.W. thanks the Swiss National Science Foundation for support under Grant No. 200021-165509, the Simons Collaborative Grant, and Aix-Marseille Universit\'e for a visiting professorship. This work is undertaken under the auspices of the `Laboratoire d'Excellence M\'ecanique et Complexit\'e' (ANR-11-LABX-0092), and the `Initiative d'Excellence' A$^{\ast}$MIDEX (ANR-11-IDEX-0001-02).
\end{acknowledgments}

\bibliography{../../bib/Wyartbibnew}

\begin{thebibliography}{24}%
\makeatletter
\providecommand \@ifxundefined [1]{%
 \@ifx{#1\undefined}
}%
\providecommand \@ifnum [1]{%
 \ifnum #1\expandafter \@firstoftwo
 \else \expandafter \@secondoftwo
 \fi
}%
\providecommand \@ifx [1]{%
 \ifx #1\expandafter \@firstoftwo
 \else \expandafter \@secondoftwo
 \fi
}%
\providecommand \natexlab [1]{#1}%
\providecommand \enquote  [1]{``#1''}%
\providecommand \bibnamefont  [1]{#1}%
\providecommand \bibfnamefont [1]{#1}%
\providecommand \citenamefont [1]{#1}%
\providecommand \href@noop [0]{\@secondoftwo}%
\providecommand \href [0]{\begingroup \@sanitize@url \@href}%
\providecommand \@href[1]{\@@startlink{#1}\@@href}%
\providecommand \@@href[1]{\endgroup#1\@@endlink}%
\providecommand \@sanitize@url [0]{\catcode `\\12\catcode `\$12\catcode
  `\&12\catcode `\#12\catcode `\^12\catcode `\_12\catcode `\%12\relax}%
\providecommand \@@startlink[1]{}%
\providecommand \@@endlink[0]{}%
\providecommand \url  [0]{\begingroup\@sanitize@url \@url }%
\providecommand \@url [1]{\endgroup\@href {#1}{\urlprefix }}%
\providecommand \urlprefix  [0]{URL }%
\providecommand \Eprint [0]{\href }%
\providecommand \doibase [0]{http://dx.doi.org/}%
\providecommand \selectlanguage [0]{\@gobble}%
\providecommand \bibinfo  [0]{\@secondoftwo}%
\providecommand \bibfield  [0]{\@secondoftwo}%
\providecommand \translation [1]{[#1]}%
\providecommand \BibitemOpen [0]{}%
\providecommand \bibitemStop [0]{}%
\providecommand \bibitemNoStop [0]{.\EOS\space}%
\providecommand \EOS [0]{\spacefactor3000\relax}%
\providecommand \BibitemShut  [1]{\csname bibitem#1\endcsname}%
\let\auto@bib@innerbib\@empty
\bibitem [{\citenamefont {Buffington}\ and\ \citenamefont
  {Montgomery}(1997)}]{Buffington97}%
  \BibitemOpen
  \bibfield  {author} {\bibinfo {author} {\bibfnamefont {J.~M.}\ \bibnamefont
  {Buffington}}\ and\ \bibinfo {author} {\bibfnamefont {D.~R.}\ \bibnamefont
  {Montgomery}},\ }\href@noop {} {\bibfield  {journal} {\bibinfo  {journal}
  {Water Resources Research}\ }\textbf {\bibinfo {volume} {33}},\ \bibinfo
  {pages} {1993} (\bibinfo {year} {1997})}\BibitemShut {NoStop}%
\bibitem [{\citenamefont {Charru}\ \emph {et~al.}(2004)\citenamefont {Charru},
  \citenamefont {Mouilleron},\ and\ \citenamefont {Eiff}}]{Charru04}%
  \BibitemOpen
  \bibfield  {author} {\bibinfo {author} {\bibfnamefont {F.}~\bibnamefont
  {Charru}}, \bibinfo {author} {\bibfnamefont {H.}~\bibnamefont {Mouilleron}},
  \ and\ \bibinfo {author} {\bibfnamefont {O.}~\bibnamefont {Eiff}},\ }\href
  {\doibase 10.1017/S0022112004001028} {\bibfield  {journal} {\bibinfo
  {journal} {Journal of Fluid Mechanics}\ }\textbf {\bibinfo {volume} {519}},\
  \bibinfo {pages} {55} (\bibinfo {year} {2004})}\BibitemShut {NoStop}%
\bibitem [{\citenamefont {Loiseleux}\ \emph {et~al.}(2005)\citenamefont
  {Loiseleux}, \citenamefont {Gondret}, \citenamefont {Rabaud},\ and\
  \citenamefont {Doppler}}]{Loiseleux05}%
  \BibitemOpen
  \bibfield  {author} {\bibinfo {author} {\bibfnamefont {T.}~\bibnamefont
  {Loiseleux}}, \bibinfo {author} {\bibfnamefont {P.}~\bibnamefont {Gondret}},
  \bibinfo {author} {\bibfnamefont {M.}~\bibnamefont {Rabaud}}, \ and\ \bibinfo
  {author} {\bibfnamefont {D.}~\bibnamefont {Doppler}},\ }\href@noop {}
  {\bibfield  {journal} {\bibinfo  {journal} {Physics of Fluids
  (1994-present)}\ }\textbf {\bibinfo {volume} {17}},\ \bibinfo {pages}
  {103304} (\bibinfo {year} {2005})}\BibitemShut {NoStop}%
\bibitem [{\citenamefont {Ouriemi}\ \emph {et~al.}(2007)\citenamefont
  {Ouriemi}, \citenamefont {Aussillous}, \citenamefont {Medale}, \citenamefont
  {Peysson},\ and\ \citenamefont {Guazzelli}}]{Ouriemi07}%
  \BibitemOpen
  \bibfield  {author} {\bibinfo {author} {\bibfnamefont {M.}~\bibnamefont
  {Ouriemi}}, \bibinfo {author} {\bibfnamefont {P.}~\bibnamefont {Aussillous}},
  \bibinfo {author} {\bibfnamefont {M.}~\bibnamefont {Medale}}, \bibinfo
  {author} {\bibfnamefont {Y.}~\bibnamefont {Peysson}}, \ and\ \bibinfo
  {author} {\bibfnamefont {{\'E}.}~\bibnamefont {Guazzelli}},\ }\href@noop {}
  {\bibfield  {journal} {\bibinfo  {journal} {Physics of Fluids}\ }\textbf
  {\bibinfo {volume} {19}},\ \bibinfo {pages} {61706} (\bibinfo {year}
  {2007})}\BibitemShut {NoStop}%
\bibitem [{\citenamefont {Derksen}(2011)}]{Derksen11}%
  \BibitemOpen
  \bibfield  {author} {\bibinfo {author} {\bibfnamefont {J.}~\bibnamefont
  {Derksen}},\ }\href@noop {} {\bibfield  {journal} {\bibinfo  {journal}
  {Physics of Fluids (1994-present)}\ }\textbf {\bibinfo {volume} {23}},\
  \bibinfo {pages} {113303} (\bibinfo {year} {2011})}\BibitemShut {NoStop}%
\bibitem [{\citenamefont {Kidanemariam}\ and\ \citenamefont
  {Uhlmann}(2014)}]{Kidanemariam14}%
  \BibitemOpen
  \bibfield  {author} {\bibinfo {author} {\bibfnamefont {A.~G.}\ \bibnamefont
  {Kidanemariam}}\ and\ \bibinfo {author} {\bibfnamefont {M.}~\bibnamefont
  {Uhlmann}},\ }\href@noop {} {\bibfield  {journal} {\bibinfo  {journal}
  {International Journal of Multiphase Flow}\ }\textbf {\bibinfo {volume}
  {67}},\ \bibinfo {pages} {174} (\bibinfo {year} {2014})}\BibitemShut
  {NoStop}%
\bibitem [{\citenamefont {Ouriemi}\ \emph {et~al.}(2009)\citenamefont
  {Ouriemi}, \citenamefont {Aussillous},\ and\ \citenamefont
  {Guazzelli}}]{Ouriemi09}%
  \BibitemOpen
  \bibfield  {author} {\bibinfo {author} {\bibfnamefont {M.}~\bibnamefont
  {Ouriemi}}, \bibinfo {author} {\bibfnamefont {P.}~\bibnamefont {Aussillous}},
  \ and\ \bibinfo {author} {\bibfnamefont {E.}~\bibnamefont {Guazzelli}},\
  }\href {\doibase 10.1017/S0022112009007915} {\bibfield  {journal} {\bibinfo
  {journal} {Journal of Fluid Mechanics}\ }\textbf {\bibinfo {volume} {636}},\
  \bibinfo {pages} {295} (\bibinfo {year} {2009})}\BibitemShut {NoStop}%
\bibitem [{\citenamefont {Bagnold}(1966)}]{Bagnold66a}%
  \BibitemOpen
  \bibfield  {author} {\bibinfo {author} {\bibfnamefont {R.}~\bibnamefont
  {Bagnold}},\ }\href@noop {} {\bibfield  {journal} {\bibinfo  {journal} {US
  Geological Survey, DOI, USA}\ } (\bibinfo {year} {1966})}\BibitemShut
  {NoStop}%
\bibitem [{\citenamefont {Chiodi}\ \emph {et~al.}(2014)\citenamefont {Chiodi},
  \citenamefont {Claudin},\ and\ \citenamefont {Andreotti}}]{Chiodi14}%
  \BibitemOpen
  \bibfield  {author} {\bibinfo {author} {\bibfnamefont {F.}~\bibnamefont
  {Chiodi}}, \bibinfo {author} {\bibfnamefont {P.}~\bibnamefont {Claudin}}, \
  and\ \bibinfo {author} {\bibfnamefont {B.}~\bibnamefont {Andreotti}},\
  }\href@noop {} {\bibfield  {journal} {\bibinfo  {journal} {Journal of Fluid
  Mechanics}\ }\textbf {\bibinfo {volume} {755}},\ \bibinfo {pages} {561}
  (\bibinfo {year} {2014})}\BibitemShut {NoStop}%
\bibitem [{\citenamefont {Houssais}\ \emph {et~al.}(2015)\citenamefont
  {Houssais}, \citenamefont {Ortiz}, \citenamefont {Durian},\ and\
  \citenamefont {Jerolmack}}]{Houssais15}%
  \BibitemOpen
  \bibfield  {author} {\bibinfo {author} {\bibfnamefont {M.}~\bibnamefont
  {Houssais}}, \bibinfo {author} {\bibfnamefont {C.~P.}\ \bibnamefont {Ortiz}},
  \bibinfo {author} {\bibfnamefont {D.~J.}\ \bibnamefont {Durian}}, \ and\
  \bibinfo {author} {\bibfnamefont {D.~J.}\ \bibnamefont {Jerolmack}},\ }\href
  {http://dx.doi.org/10.1038/ncomms7527} {\bibfield  {journal} {\bibinfo
  {journal} {Nat Commun}\ }\textbf {\bibinfo {volume} {6}} (\bibinfo {year}
  {2015})}\BibitemShut {NoStop}%
\bibitem [{\citenamefont {Yan}\ \emph {et~al.}(2016)\citenamefont {Yan},
  \citenamefont {Barizien},\ and\ \citenamefont {Wyart}}]{Yan16a}%
  \BibitemOpen
  \bibfield  {author} {\bibinfo {author} {\bibfnamefont {L.}~\bibnamefont
  {Yan}}, \bibinfo {author} {\bibfnamefont {A.}~\bibnamefont {Barizien}}, \
  and\ \bibinfo {author} {\bibfnamefont {M.}~\bibnamefont {Wyart}},\
  }\href@noop {} {\bibfield  {journal} {\bibinfo  {journal} {Physical Review
  E}\ }\textbf {\bibinfo {volume} {93}},\ \bibinfo {pages} {012903} (\bibinfo
  {year} {2016})}\BibitemShut {NoStop}%
\bibitem [{\citenamefont {Parker}\ \emph {et~al.}(2007)\citenamefont {Parker},
  \citenamefont {Wilcock}, \citenamefont {Paola}, \citenamefont {Dietrich},\
  and\ \citenamefont {Pitlick}}]{Parker07}%
  \BibitemOpen
  \bibfield  {author} {\bibinfo {author} {\bibfnamefont {G.}~\bibnamefont
  {Parker}}, \bibinfo {author} {\bibfnamefont {P.~R.}\ \bibnamefont {Wilcock}},
  \bibinfo {author} {\bibfnamefont {C.}~\bibnamefont {Paola}}, \bibinfo
  {author} {\bibfnamefont {W.~E.}\ \bibnamefont {Dietrich}}, \ and\ \bibinfo
  {author} {\bibfnamefont {J.}~\bibnamefont {Pitlick}},\ }\href {\doibase
  10.1029/2006JF000549} {\bibfield  {journal} {\bibinfo  {journal} {Journal of
  Geophysical Research: Earth Surface}\ }\textbf {\bibinfo {volume} {112}},\
  \bibinfo {pages} {n/a} (\bibinfo {year} {2007})}\BibitemShut {NoStop}%
\bibitem [{\citenamefont {Kolton}\ \emph {et~al.}(1999)\citenamefont {Kolton},
  \citenamefont {Dom\'{i}nguez},\ and\ \citenamefont
  {Gronbech-Jensen}}]{Kolton99}%
  \BibitemOpen
  \bibfield  {author} {\bibinfo {author} {\bibfnamefont {A.~B.}\ \bibnamefont
  {Kolton}}, \bibinfo {author} {\bibfnamefont {D.}~\bibnamefont
  {Dom\'{i}nguez}}, \ and\ \bibinfo {author} {\bibfnamefont {N.}~\bibnamefont
  {Gronbech-Jensen}},\ }\href {\doibase 10.1103/PhysRevLett.83.3061} {\bibfield
   {journal} {\bibinfo  {journal} {Phys. Rev. Lett.}\ }\textbf {\bibinfo
  {volume} {83}},\ \bibinfo {pages} {3061} (\bibinfo {year}
  {1999})}\BibitemShut {NoStop}%
\bibitem [{\citenamefont {Watson}\ and\ \citenamefont
  {Fisher}(1996)}]{Watson96}%
  \BibitemOpen
  \bibfield  {author} {\bibinfo {author} {\bibfnamefont {J.}~\bibnamefont
  {Watson}}\ and\ \bibinfo {author} {\bibfnamefont {D.~S.}\ \bibnamefont
  {Fisher}},\ }\href {\doibase 10.1103/PhysRevB.54.938} {\bibfield  {journal}
  {\bibinfo  {journal} {Phys. Rev. B}\ }\textbf {\bibinfo {volume} {54}},\
  \bibinfo {pages} {938} (\bibinfo {year} {1996})}\BibitemShut {NoStop}%
\bibitem [{\citenamefont {Reichhardt}\ and\ \citenamefont
  {Reichhardt}(2016)}]{Reichhardt16}%
  \BibitemOpen
  \bibfield  {author} {\bibinfo {author} {\bibfnamefont {C.}~\bibnamefont
  {Reichhardt}}\ and\ \bibinfo {author} {\bibfnamefont {C.}~\bibnamefont
  {Reichhardt}},\ }\href@noop {} {\bibfield  {journal} {\bibinfo  {journal}
  {arXiv preprint arXiv:1602.03798}\ } (\bibinfo {year} {2016})}\BibitemShut
  {NoStop}%
\bibitem [{\citenamefont {Reichhardt}\ \emph {et~al.}(2015)\citenamefont
  {Reichhardt}, \citenamefont {Ray},\ and\ \citenamefont
  {Reichhardt}}]{Reichhardt15}%
  \BibitemOpen
  \bibfield  {author} {\bibinfo {author} {\bibfnamefont {C.}~\bibnamefont
  {Reichhardt}}, \bibinfo {author} {\bibfnamefont {D.}~\bibnamefont {Ray}}, \
  and\ \bibinfo {author} {\bibfnamefont {C.~O.}\ \bibnamefont {Reichhardt}},\
  }\href@noop {} {\bibfield  {journal} {\bibinfo  {journal} {Physical review
  letters}\ }\textbf {\bibinfo {volume} {114}},\ \bibinfo {pages} {217202}
  (\bibinfo {year} {2015})}\BibitemShut {NoStop}%
\bibitem [{\citenamefont {Lajeunesse}\ \emph {et~al.}(2010)\citenamefont
  {Lajeunesse}, \citenamefont {Malverti},\ and\ \citenamefont
  {Charru}}]{Lajeunesse10}%
  \BibitemOpen
  \bibfield  {author} {\bibinfo {author} {\bibfnamefont {E.}~\bibnamefont
  {Lajeunesse}}, \bibinfo {author} {\bibfnamefont {L.}~\bibnamefont
  {Malverti}}, \ and\ \bibinfo {author} {\bibfnamefont {F.}~\bibnamefont
  {Charru}},\ }\href@noop {} {\bibfield  {journal} {\bibinfo  {journal}
  {Journal of Geophysical Research: Earth Surface (2003--2012)}\ }\textbf
  {\bibinfo {volume} {115}} (\bibinfo {year} {2010})}\BibitemShut {NoStop}%
\bibitem [{\citenamefont {Dur{\'a}n}\ \emph {et~al.}(2014)\citenamefont
  {Dur{\'a}n}, \citenamefont {Andreotti},\ and\ \citenamefont
  {Claudin}}]{Duran14}%
  \BibitemOpen
  \bibfield  {author} {\bibinfo {author} {\bibfnamefont {O.}~\bibnamefont
  {Dur{\'a}n}}, \bibinfo {author} {\bibfnamefont {B.}~\bibnamefont
  {Andreotti}}, \ and\ \bibinfo {author} {\bibfnamefont {P.}~\bibnamefont
  {Claudin}},\ }\href@noop {} {\bibfield  {journal} {\bibinfo  {journal}
  {Advances in Geosciences}\ }\textbf {\bibinfo {volume} {37}},\ \bibinfo
  {pages} {73} (\bibinfo {year} {2014})}\BibitemShut {NoStop}%
\bibitem [{Note1()}]{Note1}%
  \BibitemOpen
  \bibinfo {note} {This value of $U$ can be simply recovered by balancing the
  drag force $C_D \rho _f \pi a^2 U^2/2$ on a particle with the friction force
  on the top of the bed $4 \mu \pi a^3 (\rho _p-\rho _f) g/3$, where $C_D=[24 /
  Re_p][1+0.15 Re_p^{0.687}]$ is the Schiller-Naumann correlation for the drag
  coefficient with the particle Reynolds number defined as $Re_p= \rho _f
  aU/\eta $ and $\mu \approx 0.33$ is the friction coefficient, the value of
  which is in agreement with that found in previous work for suspensions \cite
  {Cassar05,Boyer11}. The particle Reynolds number is $Re_p=0.05$ for the
  water-Ucon mixture and $Re_p=36.10$ for pure water.}\BibitemShut {Stop}%
\bibitem [{\citenamefont {Dhar}(2006)}]{Dhar06}%
  \BibitemOpen
  \bibfield  {author} {\bibinfo {author} {\bibfnamefont {D.}~\bibnamefont
  {Dhar}},\ }\href@noop {} {\bibfield  {journal} {\bibinfo  {journal} {Physica
  A: Statistical Mechanics and its Applications}\ }\textbf {\bibinfo {volume}
  {369}},\ \bibinfo {pages} {29} (\bibinfo {year} {2006})}\BibitemShut
  {NoStop}%
\bibitem [{\citenamefont {Reichhardt}\ and\ \citenamefont
  {Olson}(2002)}]{Reichhardt02}%
  \BibitemOpen
  \bibfield  {author} {\bibinfo {author} {\bibfnamefont {C.}~\bibnamefont
  {Reichhardt}}\ and\ \bibinfo {author} {\bibfnamefont {C.}~\bibnamefont
  {Olson}},\ }\href@noop {} {\bibfield  {journal} {\bibinfo  {journal}
  {Physical review letters}\ }\textbf {\bibinfo {volume} {89}},\ \bibinfo
  {pages} {078301} (\bibinfo {year} {2002})}\BibitemShut {NoStop}%
\bibitem [{\citenamefont {Pertsinidis}\ and\ \citenamefont
  {Ling}(2008)}]{Pertsinidis08}%
  \BibitemOpen
  \bibfield  {author} {\bibinfo {author} {\bibfnamefont {A.}~\bibnamefont
  {Pertsinidis}}\ and\ \bibinfo {author} {\bibfnamefont {X.~S.}\ \bibnamefont
  {Ling}},\ }\href@noop {} {\bibfield  {journal} {\bibinfo  {journal} {Physical
  review letters}\ }\textbf {\bibinfo {volume} {100}},\ \bibinfo {pages}
  {028303} (\bibinfo {year} {2008})}\BibitemShut {NoStop}%
\bibitem [{\citenamefont {Cassar}\ \emph {et~al.}(2005)\citenamefont {Cassar},
  \citenamefont {Nicolas},\ and\ \citenamefont {Pouliquen}}]{Cassar05}%
  \BibitemOpen
  \bibfield  {author} {\bibinfo {author} {\bibfnamefont {C.}~\bibnamefont
  {Cassar}}, \bibinfo {author} {\bibfnamefont {M.}~\bibnamefont {Nicolas}}, \
  and\ \bibinfo {author} {\bibfnamefont {O.}~\bibnamefont {Pouliquen}},\
  }\href@noop {} {\bibfield  {journal} {\bibinfo  {journal} {Physics of
  Fluids}\ }\textbf {\bibinfo {volume} {17}},\ \bibinfo {pages} {103301}
  (\bibinfo {year} {2005})}\BibitemShut {NoStop}%
\bibitem [{\citenamefont {Boyer}\ \emph {et~al.}(2011)\citenamefont {Boyer},
  \citenamefont {Guazzelli},\ and\ \citenamefont {Pouliquen}}]{Boyer11}%
  \BibitemOpen
  \bibfield  {author} {\bibinfo {author} {\bibfnamefont {F.}~\bibnamefont
  {Boyer}}, \bibinfo {author} {\bibfnamefont {E.}~\bibnamefont {Guazzelli}}, \
  and\ \bibinfo {author} {\bibfnamefont {O.}~\bibnamefont {Pouliquen}},\ }\href
  {\doibase 10.1103/PhysRevLett.107.188301} {\bibfield  {journal} {\bibinfo
  {journal} {Phys. Rev. Lett.}\ }\textbf {\bibinfo {volume} {107}},\ \bibinfo
  {pages} {188301} (\bibinfo {year} {2011})}\BibitemShut {NoStop}%
\end{thebibliography}%

\end{document}